\documentclass[epj]{svjour}

\usepackage{amsmath}
\usepackage{amssymb}

\RequirePackage[T1]{fontenc}
\smartqed
\RequirePackage{graphicx}
\RequirePackage{mathptmx}
\RequirePackage{flushend}
\RequirePackage[numbers,sort&compress]{natbib}
\RequirePackage[colorlinks,citecolor=blue,urlcolor=blue,linkcolor=blue]{hyperref}

\RequirePackage{color}

\hypersetup{
  colorlinks=true,linkcolor=blue,citecolor=blue,
  filecolor=blue,urlcolor=blue,breaklinks=true
}

\begin{document}

\title{
  Simple quantum graphs proposal for quantum devices
}

\author{
  A. Drinko\inst{1} \and
  F. M. Andrade\inst{1,2} \and
  D. Bazeia\inst{3}
}

\institute{
  Programa de P\'os-Gradua\c{c}\~ao em Ci\^encias/F\'{i}sica,
  Universidade Estadual de Ponta Grossa,
  84030-900 Ponta Grossa, PR, Brazil
  \and
  Departamento de Matem\'{a}tica e Estat\'{i}stica,
  Universidade Estadual de Ponta Grossa,
  84030-900 Ponta Grossa-PR, Brazil
  \and
  Departamento de F\'{i}sica,
  Universidade Federal da Para\'{i}ba,
  58051-900 Jo\~{a}o Pessoa, PB, Brazil
}

\date{Received: date / Revised version: date}

\abstract{
The control of the quantum transport is an issue of current interest for
the construction of new devices. In this work, we investigate this
possibility in the realm of quantum graphs. The study allows the
identification of two distinct periodic quantum effects which are
related to quantum complexity, one being the identification of transport
inefficiency, and the other the presence of peaks of full transmission
inside regions of suppression of transport in some elementary
arrangements of graphs. Motivated by the power of quantum graphs, we
elaborate on the construction of simple devices, based on microwave and
optical fibers networks, and also on quantum dots, nanowires and
nanorings. The elementary devices can be used to construct composed
structures with important quantum properties, which may be used to
manipulate the quantum transport.\\
Published in Eur. Phys. J. Plus \textbf{135}, 451 (2020).
\href{https://doi.org/10.1140/epjp/s13360-020-00459-9}
{https://doi.org/10.1140/epjp/s13360-020-00459-9}
\PACS{
      {03.65.-w}{Quantum mechanics}   \and
      {03.65.Nk}{Scattering theory}
     }
}

\maketitle

\section{Introduction}
Since the pioneering work of Linus Pauling \cite{JCP.4.673.1936} in
the context of free electrons in organic molecules, quantum graphs have
been used  to describe the behavior of quantum particles in idealized
physical networks \cite{AoP.274.76.1999,AP.55.527.2006}.
In the last twenty years or so, the interest in quantum graphs
has increased importantly, mainly because of the richness of the
subject, which is related to a variety of issues in physical and
mathematical sciences
\cite{PRE.69.056205.2004,NM.3.380.2004,NL.8.4523.2008,Book.2012.Berkolaiko,
PRL.122.140503.2019,PRL.118.123901.2017,PRA.99.023841.2019,
ICEAA.2019.8879059,EMCE.2019.8871973}.
For instance, quantum graphs have been simulated experimentally in
microwave networks \cite{PRE.69.056205.2004} and it is also
possible to synthesize quantum nanowires networks with sequential
seeding of branching structures \cite{NM.3.380.2004} and also, using
conventional microfabrication facilities \cite{NL.8.4523.2008}.
Many properties and techniques related to the study of quantum graphs
and their applications can be found in \cite{Book.2012.Berkolaiko}.
Moreover, in the recent literature we noted the interesting
investigations
\cite{PRL.122.140503.2019,PRL.118.123901.2017,PRA.99.023841.2019}, in
which the authors deal with issues on graphs and simulations via
microwave networks  \cite{PRL.122.140503.2019}, in a way similar to
Ref. \cite{PRE.69.056205.2004}.
The case of networks of fibers and splitters adds another perspective
\cite{PRL.118.123901.2017} that consists in the construction of complex
active optical network of fibers and  splitters, which is further
explored in \cite{PRA.99.023841.2019} and may be nicely modelled by
graphs.
Also, quantum graphs formalism has been used to study irregular
eletromagnetic cavities \cite{ICEAA.2019.8879059} and scattering of
eletromagnetic waves propagating through networks of cables
\cite{EMCE.2019.8871973}.

Among the several possibilities to describe physical properties of
graphs, an interesting procedure concerns the Green's function approach,
which was proposed in \cite{JPA.36.545.2003} and further explored in
\cite{PR.647.1.2016,PRA.98.062107.2018}.
In this work, we rely on the Green's function approach to
investigate scattering properties of quantum graphs.
The graphs to be studied below have two leads attached to two distinct
vertices that are further connected with distinct edges.
The details of the calculations and the use of graphs with the focus on
the global transmission coefficients are all given in
\cite{PR.647.1.2016,PRA.100.62117.2019}, in which we explain and explore
carefully the concepts and techniques used to implement the calculations
that ended up with the current study.
In Ref. \cite{PRA.100.62117.2019}, in particular, we concentrated mainly
on hexagonal graphs with vertices of degree 3, with ideal leads and
edges.
We did this to focus mainly on the transmission features of simple ideal
graphs described by vertices of degree 3, in this way circumventing the
appearance of effects related with the presence of vertices of distinct
degrees.
In this work,  we deal with the two simplest regular graphs, which
contain only two vertices of degree 3.
We investigate these graphs concentrating on the possibility to develop
simple devices that engender interesting quantum effects.

The study starts with the calculation of global transmission amplitudes
of quantum graphs as a function of the wave number of the incident
signal.
The focus is on the presence of a periodic effect of transport
suppression and on the possibility to identify very narrow peaks of wave
numbers in which the transmission coefficient increases significantly
inside periodic regions of full suppression of transport.
We also examine other effects, in particular, very narrow peaks of full
transmission inside regions of full suppression.
Here, however, we concentrate on the search of peaks of full
transmission that we call peaks of constructive quantum interference,
which appear inside regions of suppression of transmission.
Evidently, the identification of those peaks allows for a diversity of
applications, including electronic transport in photosynthetic processes
that seem to offer a biological advantage
\cite{N.446.782.2007,NP.9.10.2012}, and single-molecule quantum
transport in break junctions \cite{NRP.1.381.2019}, among other
possibilities.

\begin{figure}[t!]
  \centering
  \includegraphics*[width=0.65\columnwidth]{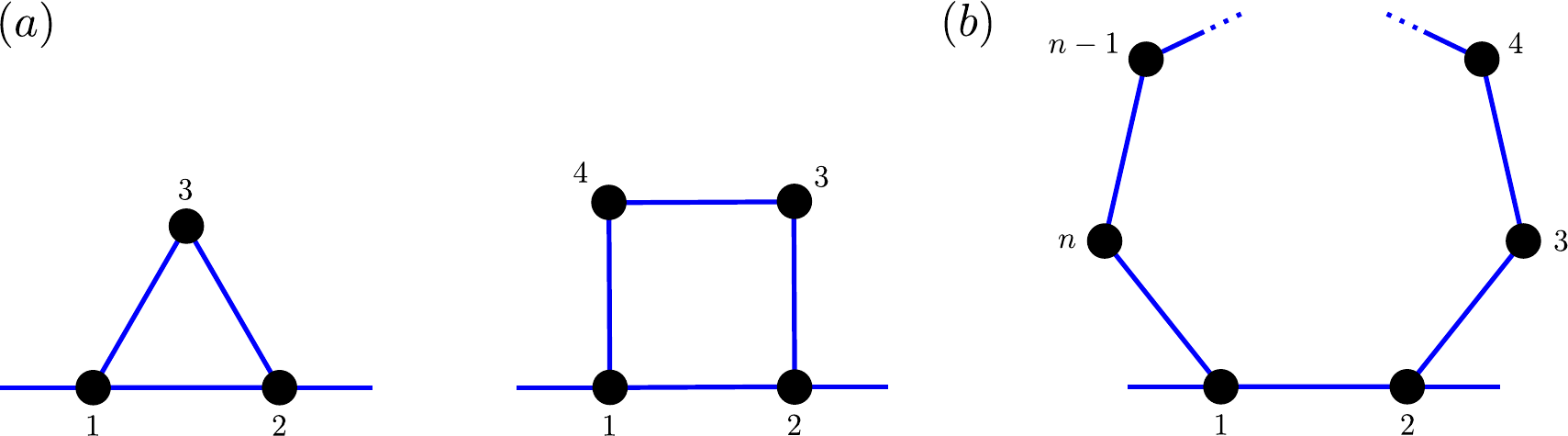}
  \caption{(Color online)
    (a) The two quantum graphs to be studied in this work,
    the regular triangle ($C_3$) with 3 vertices, and the square
    ($C_4$) with 4 vertices.
    (b) The cycle graph on $n$ vertices.}
  \label{fig:fig1}
\end{figure}

We organize the work as follows: in section \ref{sec:results} we use the Green's
function approach to obtain the transmission amplitude of some cycle
graphs, in particular for the $C_3$ and $C_4$ graphs shown in
Fig. \ref{fig:fig1}(a), and of the composition shown in Fig. \ref{fig:fig3}.
In section \ref{sec:conclusion}, we conclude the work with some comments
and suggestions of future works.

\section{Results}
\label{sec:results}

The two simple quantum graphs of interest in this work are depicted in
Fig. \ref{fig:fig1}(a).
They are constructed owing to simplicity having only two vertices of
degree 3, and they conform with the two simplest regular geometric
forms, the equilateral triangle and  the square, and we refer to them by
$C_3$ and $C_4$, that have three and four equal edges, respectively.
We consider the situation where the Schr\"odinger operator at each
edge contains no potential, since we are dealing with ideal edges:
$H=-(\hbar^2/2m)d^2/dx^2$, where $\hbar$ is the Planck constant divided
by $2\pi$ and $m$ is the mass of the particle.
Then, it is of the free particle type, so the solution has the free
particle form.
Moreover, we are considering general boundary conditions at the
vertices.
This means that at the vertex $j$ the quantum amplitudes obey the
quantum flux conservation \cite{JPA.32.595.1999}.
We can use the above information and the Green's function formalism in
\cite{PR.647.1.2016,PRA.98.062107.2018,PRA.100.62117.2019} to write down
the general solution for Green's function for the regular $C_3$  and
$C_4$ graphs with edges of the same length $\ell$ and the
external leads attached to two neighbor vertices (vertices 1 and 2 in
Fig. \ref{fig:fig1}).
The Green's function is given by
\begin{equation}
  G_{C_{n}}(x_f,x_i;k) = \frac{m}{i\hbar^2k} T_{C_{n}}(k) e^{i k(x_i+x_f)},
\end{equation}
in which $x_i$ and $x_f$ are arbitrary points in the entering and exit
leads and $n=3,4$.

For the symmetrical case in which the individual transmission amplitudes
at the vertices 1 and 2 in $C_3$ are equal, namely,
$r_2=r_1=r$, $t_2=t_1=t$, $r_3=r'$ and $t_3=t'$ by setting
$z=e^{ik\ell}$ the global transmission amplitude for $C_3$ is given by
\begin{align}\label{eq:TC3}
  T_{C_3}(k) = {}
  \frac{t^2 z}{g_{C_3}}
    \big\{[1 - (r - t)^2 z^2]t' z
    +[1 - r_3 (r - t) z^2]^2
    - (r - t)^2 t'^2 z^4\big\},
\end{align}
with
\begin{align}
  g_{C_3} = {}
    1 - r (r + 2 r') z^2 - 2 t^2 t' z^3
    + r [r r' (2 r + r') - 2 r' t^2 - r t'^2] z^4
    - (r - t)^2 (r + t)^2 (r' - t') (r' + t') z^6.
\end{align}
Likewise, for $C_4$ in which the individual transmission amplitudes are
equal at the vertices 1 and 2 and at the vertices 3 and 4, namely,
$r_1=r_2=r$, $t_1=t_2=t$, $r_4=r_3=r'$ and $t_4=t_3=t'$, the global
transmission is given by
\begin{align}
  \label{eq:TC4}
  T_{C_4}(k) = {}
    \frac{ t^2 z}{g_{C_4}}
    \left[1 - (r - t) (r' - t') z^2\right]
    \left[1 - (r - t) (r' + t') z^2\right]
    \left[1 - (r' - t') (r' + t') z^2\right],
\end{align}
with
\begin{align}
  g_{C_4} = {}
  &
    1 - (r + r')^2 z^2
    +
    2 \left[r r' (r^2 + r r' + r'^2 - t^2)
    - (r r' + t^2) t'^2\right] z^4
    \nonumber \\
  &  - \left[-r r' (r + r') + r' t^2 + r t'^2\right]^2 z^6
     + (r^2 - t^2)^2 (r'^2 - t'^2)^2 z^8.
\end{align}
The above results can be obtained standardly from Refs.
\cite{PR.647.1.2016,PRA.98.062107.2018,PRA.100.62117.2019}.
If we impose the Neumann-Kirchhoff (NK) boundary condition
\cite{Book.2012.Berkolaiko} at all vertices, we have $r=-1/3$, $t=2/3$
and $r'=0$, $t'=1$ \cite{AP.55.527.2006}.
In this case the global transmission amplitudes simplify and we can
write
\begin{align}
  T_{C_3}(k) = {}
  &
    \frac{4z(1+2z+2z^2+z^3)}{9-9z-8z^2+z^4+z^5},
    \nonumber \\
  T_{C_4}(k) = {}
  &
    \frac{4z(1-z^2)^2}{9+8z-z^6}.
\end{align}
Actually, for NK boundary condition, it is possible to write an expression
for the cycle graph on $n$ vertices, $C_n$ (see Fig. \ref{fig:fig1}(b)).
The general result is
\begin{equation}
  \label{eq:TCn}
  T_{C_{n}}(k)=
  \frac
  {4 (e^{n i k \ell}-1)(e^{i k \ell}+e^{(n-1) i k \ell})}
  {9-e^{2 i k \ell}-e^{2 (n-1) i k \ell}-8 e^{n i k \ell}+e^{2 n i k \ell}}.
\end{equation}

\begin{figure}[t!]
  \centering
\includegraphics*[width=0.5\columnwidth]{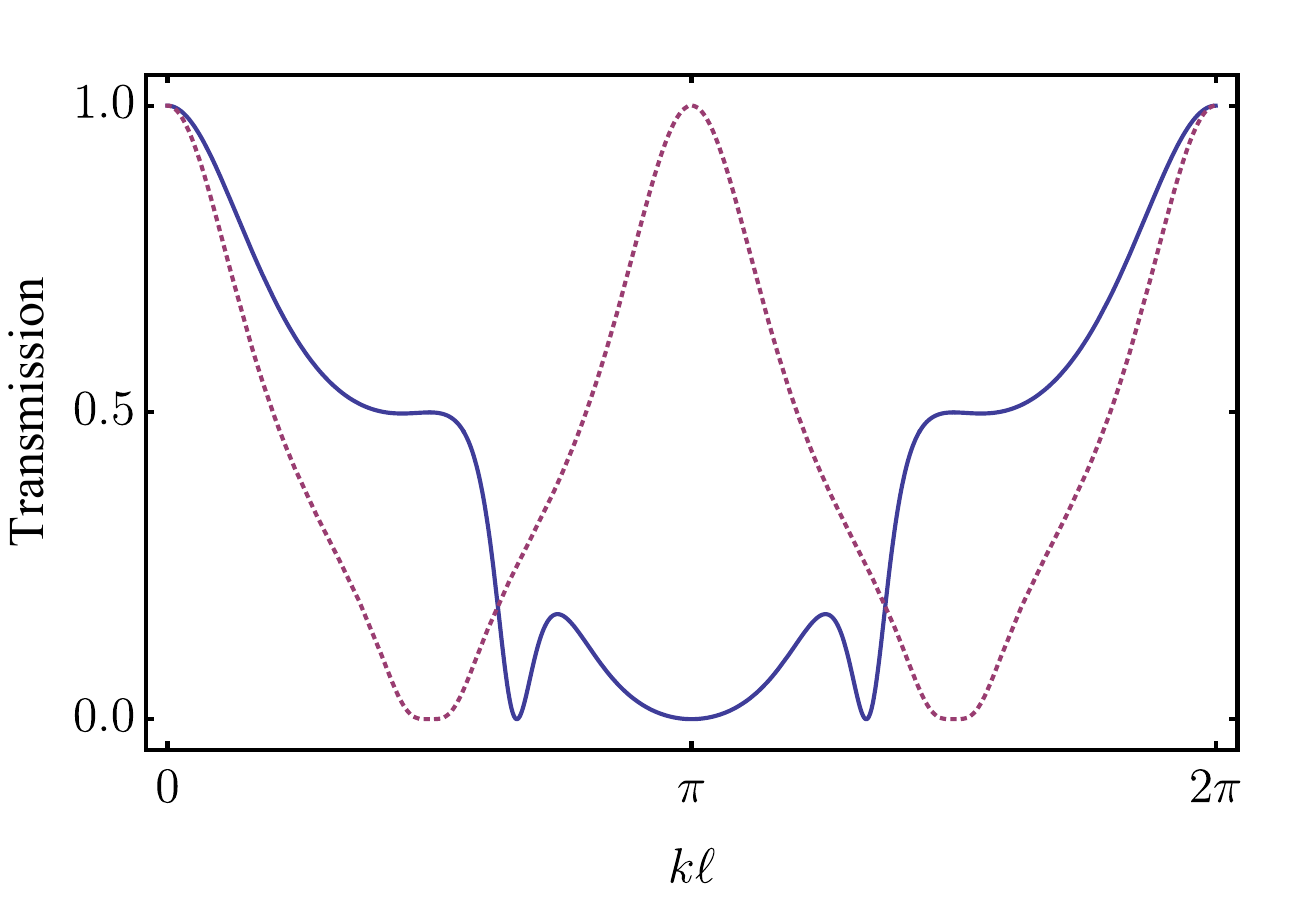}
  \caption{(Color online)
    The transmission coefficients of the two possibilities displayed in
    Fig. \ref{fig:fig1}, with the blue solid curve corresponding to the
    triangle graph and the violet dotted curve to the square graph using
    NK boundary conditions.}
  \label{fig:fig2}
\end{figure}

In Fig. \ref{fig:fig2} are displayed the transmission coefficient $|T_{C_3}(k)|^2$ and $|T_{C_4}(k)|^2$.
There, one notices that although $|T_{C_4}(k)|^2$ (violet dotted curve)
has a simpler undulating behavior, the $|T_{C_3}(k)|^2$ (blue solid
curve) on the contrary shows a richer structure.
This is an unexpected result and shows that, despite its geometric
simplicity, the triangular arrangement leads to a complex behavior that
may induce transport inefficiency.
To understand these issues more accurately, we can think of the difference
$|T_{C_3}(k)|^2-|T_{C_4}(k)|^2$: from Fig. \ref{fig:fig2} one sees that it
can be positive or negative, depending on the
wave number of the signal entering the graph, showing that the
transmission through the triangle is not always higher than the one of
the square.

To add more information on this, let us think of the calculation
in terms of the hitting time, which we interpret here as the expected
number of steps the signal has to give before leaving the graph at the
right or left side, after entering it from the left or right side.
The calculation is easier at the classical level, but since we are
interested in quantum effects, one implement it at quantum level.
The calculation of the hitting time is based on the fact that,
in the context of random walks on graphs, the concept of hitting time is
directly related to the expected number of steps to reach the vertex $f$
starting from a vertex $i$ \cite{PTRF.133.215.2005}.
This concept can be extended to the realm of quantum mechanics in the
context of quantum walks, which are the quantum version of random walks
\cite{CP.44.307.2003}.
So, the corresponding quantity for a quantum walk is the expected number
of steps to reach the quantum state in the edge $e_{n}$ starting in the
state  in the edge $e_{i}$
\cite{PLA.324.277.2004,Conference.2005.Feldman}.
Actually, quantum walks and quantum graphs are deeply related to each
other \cite{Incollection.2006.Tanner} and this relation was further
explored in \cite{PRA.84.042343.2011}.
Based on results of \cite{PRA.84.042343.2011}, it turns out that the
Green's function for a quantum graph is actually a
\textit{generating function} for all the possible walks leaving the
entrance lead and getting the exit lead in the scattering process.
In this manner, the term $z=e^{i k \ell}$ is equivalent to a time step
in the quantum walk problem. So, to extract all the paths with exactly
$m$  steps we use the \textit{step operator} \cite{PRA.84.042343.2011}
\begin{equation}
  \hat{S}_{m}=\frac{1}{m!}
  \left.\frac{\partial^m}{\partial z^m}\right|_{z=0}.
\end{equation}
In this way, the total probability for the quantum walker to leave the
entrance lead and to get the exit lead of the graph in exactly $m$ steps
in the scattering process is
\begin{equation}
  P(m)=|\hat{S}_{m} T_{C_{n}}|^{2},
\end{equation}
with $T_{C_{n}}$ given by Eqs. \eqref{eq:TC3} and \eqref{eq:TC4}.
Additionally, the probability to measure the particle for the first
time in the exit lead, regardless the number of steps, $P_{\rm out}$, is
given by
\begin{equation}
  P_{\rm  out}=\sum_{m=1}^{\infty} P(m).
\end{equation}
Then, it is possible to define a \textit{conditional hitting time}, $h$,
as \cite{PLA.324.277.2004,Conference.2005.Feldman}
\begin{equation}
  h = \frac{1}{P_{\rm out}}
  \sum_{m=1}^{\infty} m P(m).
\end{equation}
The hitting time corresponding to the two graphs under consideration,
using NK boundary conditions, are $h_{C_3}=1.91612$ and
$h_{C_4}=155/72=2.15278$, showing that the triangle is expected to be in
general more efficient in the quantum transport.
However, the effects that are captured by the transmission coefficients
of the quantum graphs make the problem more complex, inducing the
quantum transport through the triangular graph to follow the
unexpected pattern that appear in Fig. \ref{fig:fig2}, giving rise to
transport inefficiency in a large region of wave number.

\begin{figure}[t!]
  \centering
  \includegraphics*[width=0.45\columnwidth]{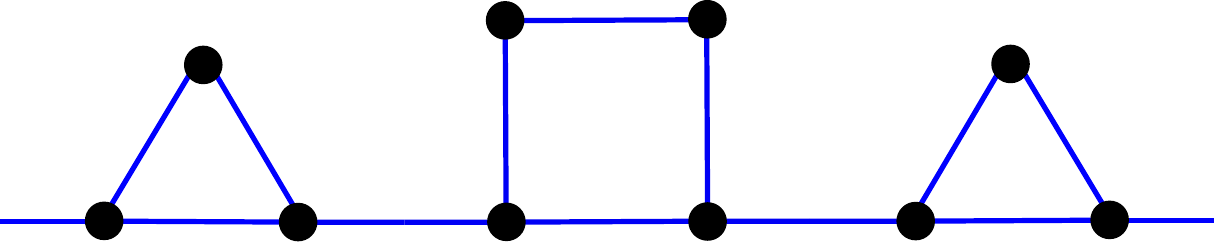}
  \caption{(Color online)
   Illustration of the series structure described by the
   triangle-square-triangle composition.}
  \label{fig:fig3}
\end{figure}

\begin{figure}[t!]
  \centering
  \includegraphics*[width=0.5\columnwidth]{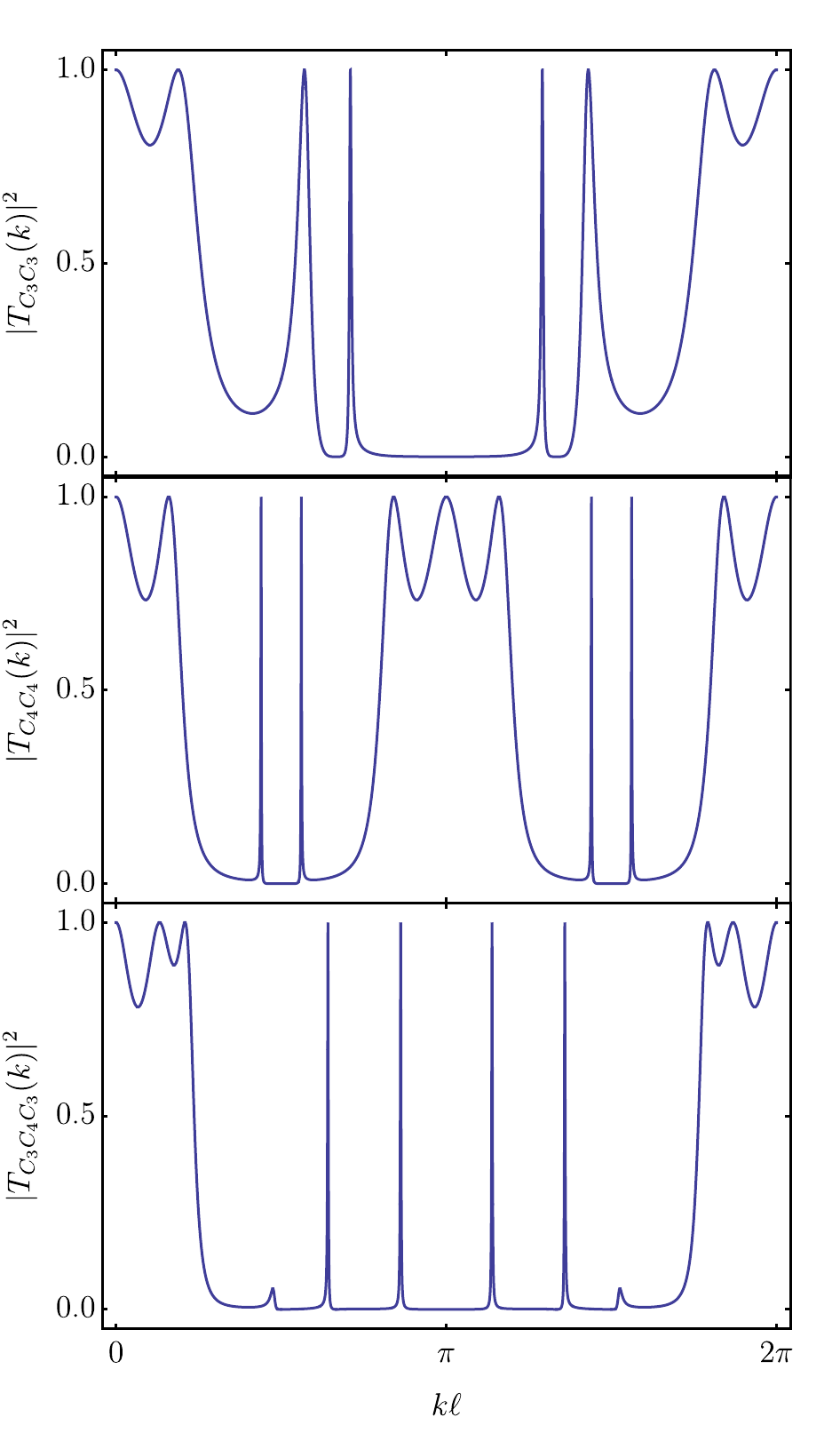}
  \caption{(Color online)
    The transmission coefficients of the series arrangements
    triangle-triangle, square-square and triangle-square-triangle,
     displayed in the top, middle and bottom panels, respectively.}
  \label{fig:fig4}
\end{figure}

The complexity of the transmission coefficients suggest that we further
explore other arrangements.
A simple possibility is to arrange the two graphs in series and in
parallel, and here we report on the interesting cases in which one adds
together the compositions triangle-triangle $(C_3C_3)$,
square-square $(C_4C_4)$, and triangle-square-triangle $(C_3C_4C_3)$;
see  Fig. \ref{fig:fig3} for an illustration of the $C_3C_4C_3$
combination.
They are series compositions of two and three elements and, in fact,
there are several parallel and series arrangements.
Here, however, we focus only on the above mentioned three cases, because
they unveil the important effects that we are searching for, in
particular, the appearance of very narrow peaks of full transmission
inside regions of complete  suppression.
The transmission coefficients of the $C_3C_3$, $C_4C_4$
and $C_3C_4C_3$ compositions are described by
$|T_{C_3 C_3}(k)|^2$, $|T_{C_4C_4}(k)|^2$, and $|T_{C_3C_4C_3}(k)|^2$,
and they  are displayed in Fig. \ref{fig:fig4}.
We see from these results that the series compositions with two
triangles and two squares present bands of full suppression and narrow
peaks of full transmission.
Similarly, the transmission coefficient of the triple composition
triangle-square-triangle has a larger band of full suppression and four
narrow peaks of full transmission.

We have studied these peaks of transmission
carefully and found that the two at the top panel are located at
$\pi \pm 0.91393$, and they have the same width $0.02091$.
The four at the middle panel are located at $\pi\pm 1.76182$ and at
$\pi\pm1.37977$ and have the same width $0.00600$.
The other four peaks at the bottom panel of Fig. \ref{fig:fig4} are
located at $\pi\pm 1.12611$, with width $0.00804$, and at
$\pi\pm0.43440$, with width $0.00812$.
The appearance of peaks of full transmission inside regions of full
suppression is another important result, that can be used to manipulate
the quantum transport.
An interesting possibility is to use compositions of the simple
triangular and square devices as filters capable of selecting signals
with wave number at some very narrow intervals, controlling the quantum
transport at the nanometric scale.
The periodicity of the effect is also of interest, since it appears in
many distinct regions of wave number having the very same
form. Moreover, the results depend on $k\ell$, so one can change $\ell$
to vary the wave number position in the filtering process.

It is of interest to remark that there are other possible studies to be
done. In particular, if one thinks of the square, for instance, we can
connect the second lead to the second neighbor vertex.
Also, we can consider the next polygon, the regular pentagon, and now we
can connect the second lead to the first or the second neighbor vertex.
Moreover, we can think of the regular hexagon, and here connect the
second lead to the first, second or third neighbor vertex, and so on
and so forth.
We have examined some distinct cases, and found no qualitative
difference with the results depicted in Fig. \ref{fig:fig4}.
Another remark of interest is that all the transmission coefficients
described in this work obey $|T(\pi+k\ell)|^2=|T(\pi-k\ell)|^2$, for
$k\ell\in[0,\pi]$.
This appears due to the time reversal symmetry of the quantum graphs
that we are examining in this work.

The two quantum graphs depicted in Fig. \ref{fig:fig1} can be thought of
as two elementary devices, so one can probe them following the lines of
Ref. \cite{PRE.69.056205.2004}, in which experimental and theoretical
results unveil how microwave networks can simulate quantum graphs.
This is an interesting possibility, and is further connected to another
very recent investigation \cite{PRL.122.140503.2019} on graphs and
simulations via microwave networks.
Another perspective of current interest is to think of considering
networks of fibers and splitters, in a way similar to the recent idea of
a {\it lasing network}, a LANER \cite{PRL.118.123901.2017} which is
constructed as a complex active optical network of fibers and splitters,
which is further explored in \cite{PRA.99.023841.2019}.
We think that the above results will foster further interest on
microwave networks
\cite{PRE.69.056205.2004,PRL.122.140503.2019,PRL.109.040402.2012} and
also, on the very recent proposal of networks of fibers and splitters
\cite{PRL.118.123901.2017,PRA.99.023841.2019}, in particular, on spatial
arrangements of graphs, with the use of more complex topologies.

Another line of research concerns the construction of quantum
devices at the nanometric scale, simulating the two quantum graphs with
QDs connected by leads.
The use of QDs for quantum computing is not new \cite{PRA.57.120.1998},
and the more recent works \cite{AA.5.117144.2015,PRL.121.257701.2018}
illustrate distinct possibilities to fabricate
compositions of two or more QDs in nanowires and in Josephson
junctions. Here we suppose that electrons in the incoming lead reach a
QD from one side and leave the device through the QD at the outgoing
lead on the other side, after interacting with the QDs via the edges
that connect them.
The flux of matter can be controlled by the chemical potentials of
electronic sources that are attached to the left and right leads.
The QD composition suggested in \cite{PRE.90.042915.2014}
will certainly display the presence of the peaks of full transmission
that we found in the series composition described in the bottom panel in
Fig. \ref{fig:fig4}.
Since QDs are mesoscopic devices, the current suggestion works to
improve the fabrication of quantum devices at the mesoscopic level, in
this sense directly contributing for the manipulation of the quantum
transport at the mesoscopic scale.

\begin{figure}[t!]
  \centering
  \includegraphics*[width=0.5\columnwidth]{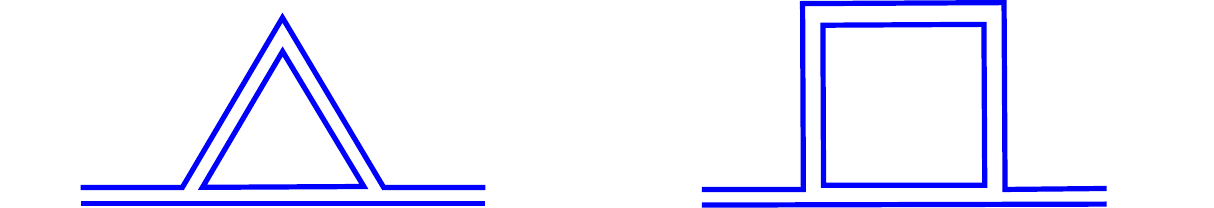}
  \caption{(Color online)
   Illustration of the two nanometric devices simulated by the
   two quantum graphs of Fig. \ref{fig:fig1}.}
  \label{fig:fig5}
\end{figure}

To implement the above suggestion experimentally, the fabrication of the
two elementary devices, the two compositions that requires two QDs
connected with two distinct edges is mandatory.
They are double QDs and we call them DQDs, and we further require series
arrangements of two and three of those DQDs.
It can then be noted that the experimental realisation may be intricate,
but we additionally observe that instead of QDs, we can use quantum
wires or nanowires, which seem to be somehow easier to be implemented
experimentally.
An interesting line of investigation may follow the experimental
implementation developed in \cite{S.303.1348.2004}, concerning the
fabrication of single-crystal nanorings of zinc oxide, formed through
spontaneous self-coiling process in the growth of polar nanobelts.
Other possibilities were accomplished, for instance, in
Refs. \cite{PRL.99.136807.2007,NP.2.826.2006}.
Similar ideas were also considered in \cite{PRL.108.076802.2012} and in
\cite{PRB.88.245417.2013}, confirming the transport inefficiency in a
ringlike disposition of wires of finite width with the addition of an
extra central branch in the ring at the nanometric scale.
The results provide another strong motivation for the study of the
series compositions that we have investigated in this work.
Since our graphs are very simple, the quantum devices are also very
simple.
The connection between ringlike quantum devices and quantum graphs is
displayed in Fig. \ref{fig:fig5}, with the identification of the two
internal arms of the nanometric devices with the internal edges of the
quantum graphs shown in Fig. \ref{fig:fig1}.
The effects of transport suppression and the presence of peaks of full
transmission can be investigated experimentally using very simple
microwave networks, similar to the one recently used in
\cite{PRL.122.140503.2019}.
Moreover, since we are dealing with ideal quantum graphs, there is no
objection to think of the triangular and square graphs as ringlike
structures, with the external leads being attached radially, in the
first case forming an angle of $2\pi/3$, and in the second an angle of
$\pi/2$.
The experimental construction may follow \cite{S.303.1348.2004,NP.2.826.2006,PRL.99.136807.2007}
and take advantage of the relation required to inscribe equilateral
triangles and squares in circles, keeping the (integer) ratios $2/1$ and
$3/1$ between the length of the two internal edges in the triangular and
square cases, respectively.

\section{Conclusion}
\label{sec:conclusion}

In this work, we use the Green's function approach to investigate the
transmission coefficient of simple quantum graphs and some of their
compositions, in particular the composition shown in
Fig. \ref{fig:fig3}.
It is interesting to note that the transmission coefficient displayed
in the bottom panel in Fig. \ref{fig:fig4} shows a large region of
transport suppression, besides the presence of some peaks of full
transmission.
This kind of behaviour suggests that this structure can be used as a
filter for the quantum transport at the nanometric scale.

From the theoretical point of view, the current work motivates several
distinct possibilities, an interesting one being the study of graphs
with other topologies in the presence of more realistic edges and
vertices; see Ref. \cite{PR.647.1.2016} for more details on how to
include different effects on edges and nontrivial boundary conditions on
vertices.
We can also add appropriate external magnetic fields, which will break
the time reversal symmetry and add novel effects in the important case
of quantum transport.
Another perspective of current interest is to think of the elementary
quantum graphs and their series and/or parallel arrangements as
\textit{molecules}.
In this case, we can use them to describe quantum transport and
interference with the help of mechanical and electrical break junctions,
as recently reviewed in Ref. \cite{NRP.1.381.2019,NRP.1.211.2019}.
A particularly important realization may occur with the use of hexagonal
arrangements of graphs, for instance, since they can model graphene and
other ringlike configurations of current interest.
This is now under examination, and we expect to report on the issue in
the near future.

\section*{Acknowledgements}
This work was partially supported by the Brazilian agencies Conselho
Nacional de Desenvolvimento Cient\'\i fico e Te\-cnol\'ogico (CNPq),
Funda\c{c}\~{a}o Arauc\'{a}ria (FAPPR, Grant 09/2016),
Instituto Nacional de Ci\^{e}ncia e Tecnologia de Informa\c{c}\~{a}o
Qu\^{a}ntica (INCT-IQ), and
Paraiba State Research Foundation (FAPESQ-PB, Grant 0015/2019).
It was also financed by the Co\-or\-dena\c{c}\~{a}o de
Aperfei\c{c}oamento de Pessoal de N\'{i}vel Superior (CAPES, Finance
Code 001).
FMA and DB also acknowledge CNPq Grants
313274/2017-7 (FMA), 434134/2018-0 (FMA),
306614/2014-6 (DB) and 404913/2018-0 (DB).

\bibliographystyle{spphys}

\end{document}